\documentclass[sigconf]{acmart}
\acmConference[EASE 2025]{The 29th International Conference on Evaluation and Assessment in Software Engineering}{17–20 June, 2025}{Istanbul, Turkey}
\AtBeginDocument{%
  }

\setcopyright{acmlicensed}
\copyrightyear{2025}
\acmYear{2025}
\acmDOI{XXXXXXX.XXXXXXX}

\acmISBN{978-1-4503-XXXX-X/2018/06}

\usepackage{booktabs}
\usepackage{subcaption}
\usepackage{graphicx}
\newcommand{\todo}[1]{}
\renewcommand{\todo}[1]{{\color{red} TODO: {#1}}}
\usepackage{comment}

\usepackage{amssymb}
\usepackage{multirow}
\usepackage{lscape}
\usepackage{array, multirow, bigdelim, makecell}
\usepackage{fontawesome5}

\usepackage[skins]{tcolorbox} 
\usepackage{cuted}
\usepackage{enumitem}
\usepackage{algorithm2e}
\RestyleAlgo{ruled}
\usepackage{longtable}
\usepackage{afterpage} 
\usepackage{pdfpages}
\usepackage{url} 

\begin{document}
\title{How Are We Doing With Using AI-Based Programming Assistants For Privacy-Related Code Generation? The Developers' Experience}

\author{Kashumi Madampe}
\email{kashumi.madampe@monash.edu}
\orcid{0000-0003-1363-8786}
\affiliation{%
  \institution{Monash University}
  \city{Melbourne}
  \state{Victoria}
  \country{Australia}
}

\author{John Grundy}
\email{john.grundy@monash.edu}
\orcid{0000-0003-4928-7076}
\affiliation{%
  \institution{Monash University}
  \city{Melbourne}
  \state{Victoria}
  \country{Australia}
}

\author{Nalin Arachchilage}
\email{nalin.arachchilage@rmit.edu.au}
\orcid{0000-0002-0059-0376}
\affiliation{%
  \institution{RMIT University}
  \city{Melbourne}
  \state{Victoria}
  \country{Australia}
}

\begin{abstract}
With generative AI becoming widespread, the existence of AI-based programming assistants for developers is no surprise. Developers increasingly use them for their work, including generating code to fulfil the data protection requirements (privacy) of the apps they build. 
We wanted to know if the reality is the same as expectations of AI-based programming assistants when trying to fulfil software privacy requirements, and the challenges developers face when using AI-based programming assistants and how these can be improved.
To this end,  we conducted a survey with 51 professional developers worldwide.
We found that AI-based programming assistants need to be improved in order for developers to better trust them with generating code that ensures privacy. In this paper, we provide some recommendations including model and system-level improvements and some key further research directions to improve AI-based programming assistants for developing secure code.
\end{abstract}




\maketitle

\section{Introduction}

Given the capacity of generating human-like text using the large corpus of data the large language models (LLMs) were trained on, the popularity of using LLMs in applications of different domains is rising exponentially.  Codex -- OpenAI’s LLM model used to generate code in GitHub’s Copilot  -- claims that it is ``proficient in more than a dozen programming languages, Codex can now interpret simple commands in natural language and execute them on the user’s behalf—making it possible to build a natural language interface to existing applications''\footnote{https://openai.com/index/openai-codex/}. Similar to Copilot, other LLM-driven code generators (AI-based programming assistants) such as Amazon Q Developer (Amazon Code Whisperer), Tabnine, Kite etc., are widely available for the developer community to use. These AI-based programming assistants claim that they enhance productivity, which is key in software development\footnote{https://aws.amazon.com/q/developer/, https://github.com/features/copilot} and some research has found the same results \cite{Imai2022IsStudy, Peng2023TheCopilot}.
However, AI-based programming assistants struggle with complex tasks \cite{Tian2023IsIt}, the accuracy of the code of these AI-based programming assistants depends on the programming language and the task \cite{Nguyen2022AnSuggestions, Yetistiren2022AssessingGeneration, MoradiDakhel2023GitHubLiability}, with longer details resulting in poor results \cite{Camara2023OnUML}, and developers may spend more time reviewing the code that AI-based programming assistants produced \cite{Bird2023TakingCopilot}. The developers also face challenges with understanding, editing, and debugging the code generated by the tools \cite{Vaithilingam2022ExpectationModels, Liang2023UnderstandingAssistants}, resulting in inefficiency \cite{Mozannar2022ReadingProgramming}.

GDPR defines data protection (``privacy" as used in this paper), as ``keeping data safe from unauthorised access'' \cite{GeneralDataProtectionRegulationArt.Definitions}. Privacy engineering encourages embedding privacy into systems. For example, designing clear privacy controls (on the user-facing side), determining the best methods for anonymisation, and inspecting code before deployment to assess privacy risks. Proactive measures of privacy engineering include data minimisation and retention \cite{Senarath2019ASystems}. 

For a business to leverage the data it has while ensuring the privacy of its users' personal data, privacy-enhancing technologies (PETs) can be used. PETs are important due to the necessity of following data protection laws such as GDPR and CCPA, requirements of testing data by third parties, and protecting the business from tarnishing its reputation from privacy breaches.
However, developers struggle with embedding privacy into software systems \cite{Anthonysamy2017PrivacyFuture, Farhadi2019ComplianceRequirements, Notario2015PRIPARE:Methodology}. They find it difficult to relate privacy requirements to privacy techniques, they see privacy in contradiction with system requirements, developers personal and peers' opinions affect how they design, and they lack privacy knowledge \cite{Senarath2018c}. 
When developers are asked to consider privacy in application designs, developers intuitively focus on technical aspects, but not user privacy expectations. Developers are also observed to have a reduced level of privacy expectations compared to users \cite{Senarath2018b}.
This leads to several critical issues. For example, permission-related issues frequently happen in Android apps \cite{Scoccia2019PermissionStudy}. 
Having tools such as AI-based programming assistants comes in handy for developers in all the above-mentioned cases. Therefore, the developers may want to use AI-based programming assistants to generate code to meet their software's privacy requirements. 

Nevertheless, the widespread uptake of AI-based programming assistants to implement systems -- including data privacy-related code -- raises the potential of their use for more secure code construction but, on the other hand, also introduces unexpected and unintended code vulnerabilities \cite{gupta2023chatgpt}. To better understand the practical experience of using AI-based programming assistants for privacy-related code generation, we carried out a survey
\footnote{Approved by Monash Human Research Ethics Committee. Project ID: 37904.} 
of 51 software developers worldwide. We asked the developers about their experience with AI-based programming assistants for privacy-related code generation. Our research yielded knowledge and insights on AI-based programming assistants, including the difference between the expectation and reality of AI-based programming vis-à-vis meeting privacy-related requirements in AI-generated code and the critical need for improvements in the AI-based programming assistants for developers to trust them with generating code for ensuring privacy. 

\section{Research Questions}

We wanted to answer the following research questions in this work:

\begin{itemize}
   \item [RQ1] \textbf{Is the reality the same as the expectation of AI-based programming assistants when it comes to ensuring privacy requirements are met in the code? } 
     Privacy is a crucial non-functional requirement in software development. We deep-dived into this aspect on the subject of the code generated by AI-based programming assistants. As the first step, we evaluated if the reality is similar or different to developers' expectations (Section \ref{expectations_reality}).
    \item [RQ2] \textbf{ What are the challenges with generating AI-based programming assistants to ensure meeting privacy requirements? }
    We wanted to explore the challenges concerning privacy-ensuring code as identifying the challenges is essential for improving the AI-based programming assistants (Section \ref{challenges}).
    \item [RQ3] \textbf{How can AI-based programming assistants be improved? }
    Developers are the users of AI-based programming assistants. It is critical to get their feedback on improvements to these  programming assistants based on their experiences. We used this question to investigate potential improvements in AI-based programming assistants (Section \ref{improvements}).
\end{itemize}

\section{Our Approach -- The Developer Survey}

\textbf{Survey Questionnaire Development. }
To answer the research questions, we conducted a developer survey. We designed our survey with both open-ended and closed-ended questions with a mapping against our research questions. 
For example, to answer RQ1, we used both open-ended and closed-ended questions. For the complete survey questionnaire (this includes questions beyond the scope of this paper), please check our online replication package 
\footnote{https://github.com/kashumi-m/AICodeGensPrivacy}. 
The questions used to collect data on company size, developers' roles, and employment status reported in this paper were adapted from the Stack Overflow developer survey\footnote{https://survey.stackoverflow.co/2022}. After we developed the questionnaire, we piloted it with the help of two postdoctoral researchers who worked in the software industry in the past at the 
HumaniSE Lab at Monash University.
They gave feedback on completion time, and the flow of the questions. Upon their feedback, we moved the demographic questions to the end of the questionnaire, which were at the beginning earlier.

\textbf{Data Collection and Analysis. }
After finalising the questionnaire, we hosted the survey on \textit{Qualtrics} and recruited developers who work in the IT industry via the recruitment platform \textit{Prolific}. We spent 6 Sterling Pounds on each participant. The participant spent around 15-20 minutes to fill out the questionnaire. We recruited 51 participants and we were able to generate rich insights from the data we collected. The details of the participants are given in the next section.
To analyse the data we collected through open-ended (qualitative) and closed-ended (quantitative) questions, we used open coding and constant comparison \cite{Strauss1998} -- interpreting the data using small chunks of words (codes) and categorising them by comparison, and with descriptive statistics. 

\section{Findings}
\subsection{Survey Participants}
\label{context}
\subsubsection{Demographics of the Participants}
The majority of the developers who participated in our study were men (76.47\%; n=39) and the participants resided around the world. 
Most participants were from companies with 101–500 employees (17.65\%; n=9).
The majority were from the IT industry (78.43\%; n=40).
Most of our participants were full-time employees (88.84\%; n=45) and the majority worked hybrid -- some remote and some in-person (52.94\%; n=27). 
The participants had a median of 9 years of experience in the software industry, 7 years of professional coding experience, and 6 years of agile experience.

\subsubsection{Who Used AI-based Programming Assistants the Most and Who Used Them the Least}
While our participants played multiple roles in their jobs, 
the majority were full-stack developers (37.25\%; n=19). The majority of the developers used AI-based programming assistants 1–2 times a day (54.90\%; n=28). From that, most were full-stack developers (21.57\%; n=11). Out of the participants who never used AI-based programming assistants (19.61\%; n=10) were the back-end developers (7.84\%; n=4). ChatGPT was the most commonly used (61.54\%; n=32) AI-based programming assistant by our developers. 

\subsection{RQ1 -- Expectation vs Reality of AI-based Programming Assistants for Privacy-related Code Generation}
\label{expectations_reality}
\begin{figure}[b]
    \centering
    \includegraphics[scale=0.4]{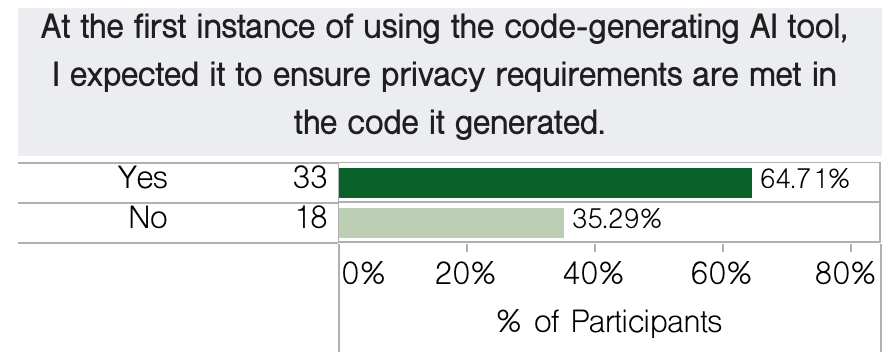}
    \caption{Developers \textit{expected} that the AI-based programming assistant will ensure meeting the privacy requirements.}
    \label{fig:expectation}
\end{figure}

\begin{figure}[b]
    \centering
    \includegraphics[scale=0.4]{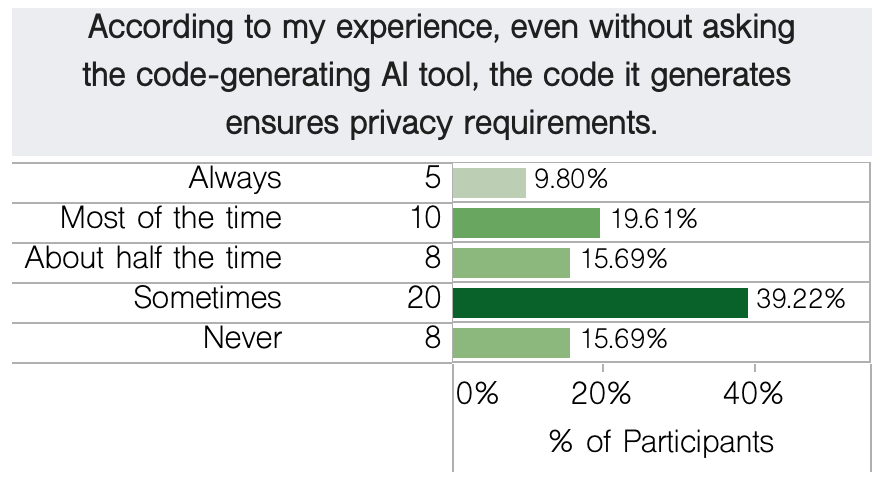}
    \caption{Developers found that AI-based programming assistants \textit{sometimes} generate code meeting privacy requirements.}
    \label{fig:reality-automated_code}
\end{figure}

\begin{figure}[b]
    \centering
    \includegraphics[scale=0.4]{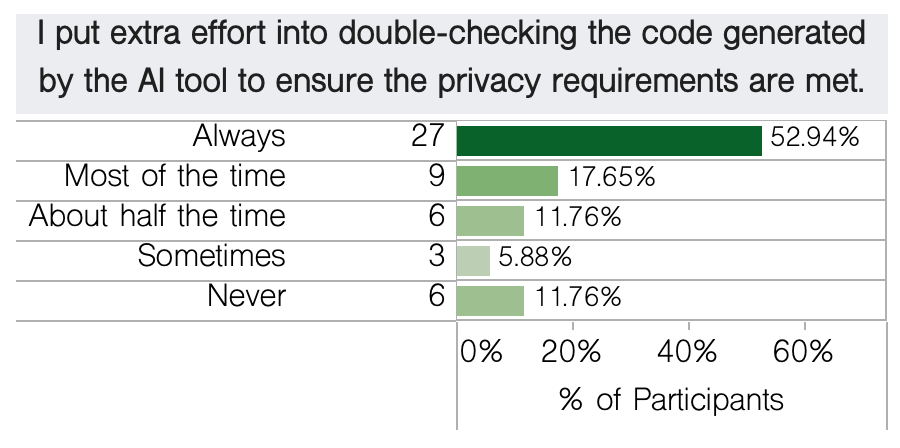}
    \caption{Developers \textit{always} put extra effort to double-check the code generated by the AI tool to ensure the privacy requirements were met.}
    \label{fig:reality_double_check}
\end{figure}

\begin{figure}[b]
    \centering
    \includegraphics[scale=0.4]{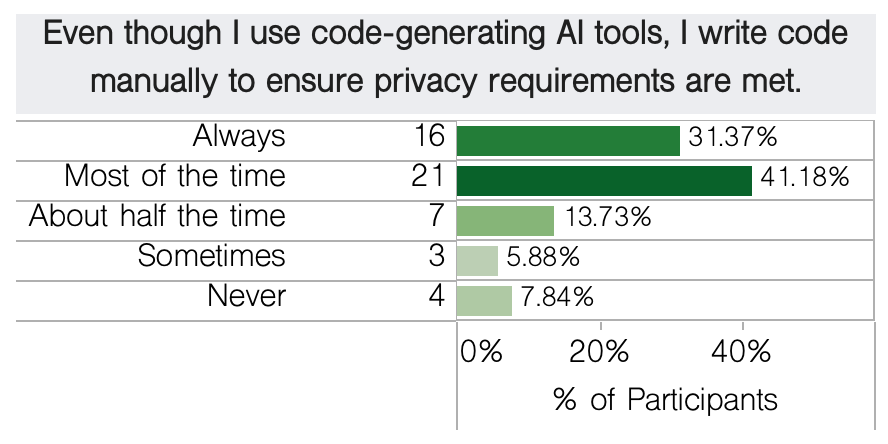}
    \caption{\textit{Most of the time}, the majority of the developers wrote code manually to ensure privacy requirements were met.}
    \label{fig:reality_manually_code}
\end{figure}

\begin{figure}[b]
    \centering
    \includegraphics[scale=0.4]{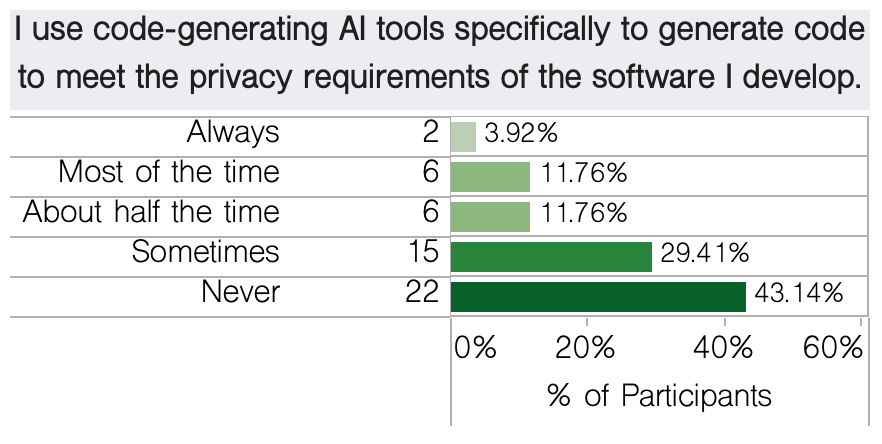}
    \caption{Majority of the developers \textit{never} used AI-based programming assistants to generate code to meet privacy requirements.}
    \label{fig:reality_never_use}
\end{figure}

When first using an AI-based programming assistant (tool), the majority of the developers (64.71\%; n=33) expected that the tool will generate code ensuring that key data privacy requirements are met (Fig. \ref{fig:expectation}). However, as they moved forward with using the tool, reality became somewhat different from what they had expected.
The majority (39.22\%; n=20) found that AI-based programming assistants \textit{sometimes} generate code ensuring privacy requirements without requesting it to meet the requirement (Fig. \ref{fig:reality-automated_code}).  However, they had to \textit{always} (52.94\%; n=27) put extra effort into double-checking the generated code to see if privacy requirements were met (Fig. \ref{fig:reality_double_check}). Therefore, unsurprisingly, a large number of the developers (43.14\%; n=22) \textit{never} used AI-based programming assistants to generate code for privacy requirements (Fig. \ref{fig:reality_never_use}), and many of them (41.18\%, n=21) wrote code manually \textit{most of the time} to meet privacy requirements (Fig. \ref{fig:reality_manually_code}), even though they used AI-based programming assistants during software development.

\subsection{RQ2 -- Key Challenges in Generating Code Ensuring Privacy using AI-based Programming Assistants}
\label{challenges}
We dug deeper into insights that our surveyed developers shared 
Key issues reported included not generating privacy-preserving code, poor output quality, distrust and use of alternative approaches to ensure data privacy, and extra effort needed.

\subsubsection{Unfulfilled task - not generating privacy-preserving code (19.61\%; n=10)}
Our developers mentioned that current AI-based programming assistants lack deep comprehension of and code generation meeting diverse privacy regulations. They said that AI-based programming assistants may not be able to keep up with changing privacy policies, they may not fully understand privacy rules or specific needs. Hence, the AI-based programming assistants may struggle to implement customised privacy controls:
\textit{``When I use AI-code generators to ensure privacy in the software I create, I face some challenges. One challenge is that these tools may not fully understand privacy rules or the specific needs of the software. Privacy requirements can be complex, and the tools may struggle to implement customized privacy controls'' -- P39 (Developer, full-stack, 6 years exp.)}

\subsubsection{Poor output quality (9.80\%; n=5)}
Many developers experienced generation of buggy code, code which had discrepancies, and ended up with code that did not work properly. They also said that they found that the coding style is different from humans:
\textit{ ``I have encountered some challenges thus far - incorporating nuances of different privacy regulations across jurisdictions (different states and territories across Australia) can lead to discrepancies in the generated code'' -- P28 (Developer, back-end, desktop or enterprise applications, 20 years exp.)}

\subsubsection{Lack of trustworthiness - Distrust and use of alternatives to ensure privacy (64.71\%; n=33)}
Most of our surveyed developers said they do not trust AI-based programming assistants with sensitive data, and it is crucial to use them for this task very cautiously. They said that AI is not self-sufficient for privacy requirements, therefore, human intervention is necessary. Therefore, apart from writing manual code they also use other means such as generating code using example data and then amending it or relying on more reliable sources or colleagues:
\textit{``At the moment I still don't blindly trust using these systems to write sensitive parts of the software, I prefer to rely on more reliable sources or the experience of colleagues who know better than me how to manage confidential information in the software we produce'' -- P19 (Developer, front-end, 4 years exp.)}

\subsubsection{Extra effort required (35.29\%; n=18)}
Many developers said they had to double-check privacy-related code more carefully than other generated code and that the amount of time they had to spend on quality assurance tasks was increased. 
``Needing to double check to make sure privacy needs are meet. In doing so, sometimes it takes up valuable time which can be used for other productivity'' -- P39 (Developer, full-stack, 6 years exp.)

However, many of these developers did believe that it was their responsibility to ensure that privacy requirements are met. Therefore, it is necessary to thoroughly review the code and perform necessities such as impact assessments:
\textit{``...developers still bear the responsibility for ensuring privacy requirements are met. It is essential to review and test the generated code thoroughly, conduct privacy impact assessments, and have appropriate mechanisms in place to address any privacy concerns or issues that may arise'' -- P1 (Developer, full-stack, 3 years exp.)}

\subsection{RQ3 -- Suggestions for Improvements to AI-based Programming Assistants for Privacy-related Code Generation}
\label{improvements}
Our developers shared what they thought could be improved in current AI-based programming assistants to produce more useful and secure privacy-preserving code. The suggestions included understanding of diverse data privacy laws, self testing, explainable output, improving transparency, following standards, compliance, ease of use, recommendations for data protection, and improving interpretability.

\subsubsection{Understand Diverse Data Privacy Laws (24.75\%; n=14)}
The most common suggestion from our developers is to train the AI to better ensure data protection in generated code. They mentioned that regular and more complete training and testing are necessary. They said that this training needs to be done using reliable and verified code: 
   \textit{ ``Perhaps these tools should be trained on reliable and verified code to make them generate code of a high enough quality to be used in corporate contexts when it comes to meeting privacy standards'' -- P17 (Developer, full-stack, 3 years exp.).}
Further, they mentioned that AI needs to have bigger contextual thinking and robust algorithms that understand data privacy laws in different countries and regions which are significantly different \cite{baumer2004internet}. 

\subsubsection{Self Testing (7.84\%; n=4)}
Even experienced developers make mistakes with privacy-related code \cite{oliveira2018api}. Our developers suggested AI-based programming assistants should be more responsible and “self-test” the code provided. This would include checking that all privacy requirements are met by it before providing developers with the final code. This is important to generate better code which again will increase the safety of the code and usage of the tools:
    \textit{``I think the generated code should undergo a test after generating to ensure the privacy requirements are met'' -- P39 (Developer, full-stack, 6 years exp.)}

\subsubsection{Explainable Output (3.92\%; n=2)}
Current AI-based programming assistants fail to clearly explain rationale behind code produced. They should generate code with attention mechanisms and comments that explain to privacy-related code produced, including the list of the resources it has used:
    \textit{``Techniques such as attention mechanisms or explainable AI methods can help developers identify potential privacy risks or vulnerabilities in the generated code'' -- P45 (Developer, back-end, front-end, mobile, 12 years exp.)}

\subsubsection{Improve Transparency (11.76\%; n=6)}
Our developers said that the transparency of the owning company’s privacy policies and practices including data usage need to be improved. This is highly beneficial for AI code gen SaaS companies as building trust with the users will help in increasing the usage:
    \textit{``We need better guarantee about the security of the tool itself, and the use of data by the company that owns it'' -- P25 (Developer, back-end, 10 years exp.)}

\subsubsection{Follow Standards (7.84\%; n=4)}
From a governance point of view, some developers said that it is necessary to define a set of privacy requirements for the AI to follow. They said that following a set of lawful guidelines is necessary \cite{gasiba2020awareness}:
    ``The world needs to decide on what privacy requirements we are supposed to follow for current AI to work for it'' -- P12 (Developer, back-end, 3 years exp.)

\subsubsection{Compliance (1.96\%; n=1)}
One important suggestion was that AI-based programming assistants should show that they meet various compliance requirements \cite{klymenko2022understanding}. The companies should audit their AI tools by an external party to certify them. Such certifications will bring better trustworthiness to the tools and hence to the code they generate and the software solutions they help to produce:
   \textit{ ``Unfortunately there is no way from AI tool user to ensure at 100\% that privacy requirements are met by AI tool, so users needs to trust the company that releases the AI tool. One way for having a certain level of "privacy requirement ensuring" is to audit the AI tool by an external company, who certifies that privacy requirements are met'' -- P6 (Developer, full-stack, 10 years exp.)}

\subsubsection{Ease of Use (13.74\%; n=7)}
AI code gen tools should provide more user control over the level of compliance with specific privacy regulations, kind of code generated, use of APIs in the code, etc \cite{hussain2020enterprise}. The developers said that AI-based programming assistants should generate code that is abstract, extensible, and also sample code:
    \textit{``Generate not that specific code, and then the programmer can extend it'' -- P44 (Developer, front-end, 6 years exp.)}

\subsubsection{Recommendations for Data Protection (17.65\%; n=9)}
AI-based programming assistants need to be able to detect that user-sensitive data are being processed within the target code and be able to handle them properly \cite{gupta2023chatgpt}. Our developers said that the AI code gens should notify and warn the user (i.e., the developer) about possible risks and provide recommendations to the users about ensuring data protection. They further said that AI-based programming assistants should introduce privacy verification checks. But, one developer said that human errors are the main problem source, so the developers need to be well aware of data privacy requirements:
   \textit{ ``Well, it could detect that user sensitive data are being processed with the code and ensure that they are handled properly (e.g. inserting into the database). It could also notify the programmer and give him/her recommendations about how to use the generated code to ensure sensitive data safety'' -- P34 (Developer, mobile, 10 years exp.)}

\subsubsection{Improve Interpretability (3.92\%; n=2)}
As an UI improvement, a few developers suggested improving interpretability including visual aspects such as highlighting areas where privacy requirements are violated in generated code:
    \textit{``The tools ($<$AI code gens$>$) can highlight areas that it thinks can violate privacy requirements. Otherwise, it should be the duty of the programmer to ensure privacy requirements are met'' -- P41 (Developer, full-stack, 30 years exp.)}

\section{Discussion}

Our surveyed developers said that AI-based programming assistants are still not mature enough to be used extensively for privacy-preserving code. AI-based programming assistants do not generate code compliant with all privacy requirements and laws. Therefore, relying completely on AI-based programming assistants to generate privacy-related code is not a wise decision. Our developers had concerns about using AI-based programming assistants because of the distrust they have regarding both their training data and their provided code context data. Developers want more information about generated code explanation, source, and compliance to different privacy-related requirements. They recommend not to use real data when generating code for the app in production, but rather use examples to generate code. Key priorities for improvement include improving model training and introducing robust algorithms that understand data privacy laws, self-testing the generated code, generating explainable outputs, improving transparency, following standards, having compliance, improving ease of use, recommendations for privacy-preserving requirements, and improving interpretability.

\subsection{Implications for Practice}
\textbf{Model and system-level improvements. }
Better model training and introducing robust algorithms that understand multiple data privacy laws, code that follows standards, generating more explainable output, and running self-tests before providing code to developers. Calibrating the AI-based programming assistants against industry-standard trustworthy AI checks will help mitigate unnecessary threats and disadvantages for AI-based programming assistants to survive in the market.

\textbf{Human-AI interaction improvements. }
Ease of use in prompting and including generated code, interpretability, highlighting sensitive data, highlighting non-compliant code, and notifying/providing recommendations for privacy approaches can enhance human-AI interaction. 

\textbf{Organisational improvements. }
Organisations owning AI-based programming assistants need to provide better compliance and improve transparency about their models and how data is gathered, processed and models trained. This is extremely important to gain trust from users, both code developers and code end users.  

\subsection{Limitations and Implications for Research}
AI-based programming assistants might have changed in various ways, hence the experience with them could be different from what our participants experienced, and what we experienced. Therefore, as a next step, researchers may continue conducting research on these tools/ replicating our study to see the progress of the AI-based programming assistants (if they have made any). 
How the AI-based programming assistants behave may differ from case to case. In this paper, we have focused on generating code for privacy. However, the results may be different in other cases, for example when generating code for other functional/non-functional requirements. In the future, researchers may consider conducting research to see how AI-based programming assistants behave across various use cases.
The number of participants who participated in our study was limited, and we did not have the chance to ask follow-up questions. To gain a more in-depth understanding of the developers' experience with AI-based programming assistants, further research can be conducted using other data collection methods such as interviews.
Conducting experiments with professional developers will help in further understanding their expectations and their interactions with AI-based programming assistants. In future, experiments with professional developers can be conducted to better understand human (developer)-AI interaction.

\subsection{Related Work}

Even though the experiment conducted by Tian et al. \cite{Tian2023IsIt} was not related to privacy, they used medium and hard problems on Leetcode. The authors found that ChatGPT struggles to generate code for new and unseen problems.		
Delile et al. \cite{Delile2023EvaluatingCompete} state that Stack Overflow slightly outperforms ChatGPT when answering privacy questions, but ChatGPT generated code can be used as alternatives.	
Pearce et al. \cite{Pearce2022AsleepContributions} found cases where insecure code was generated by GitHub Copilot.	
According to Tang et al. \cite{Tang2023AnCode},	when GitHub Copilot generated large chunks of code, the participants in their study stated that the code was cumbersome and it required significant effort to understand the meaning of generated code.	
As per Bird et al.'s work \cite{Bird2023TakingCopilot}, the first-time users of GitHub Copilot spend more time reviewing the code generated by the AI code generator than writing code.The authors also mention in their work that he early users of GitHub Copilot reported that there was a risk of revealing secrets such as API keys and passwords or suggest inappropriate test. Further, the authors state that this was fixed at the technical preview of GitHub Copilot. However, according to our findings, the distrust of AI-based programming assistants and use of alternatives to ensure privacy still exist.
Sarkar et al.'s work \cite{Sarkar2022WhatIntelligence} mentions that developers' programming shifts from writing code to checking and doing unfamiliar debugging.	
Cheng et al. \cite{Cheng2022ItTools} recommend that it is necessary to assure users about privacy and confidentiality. According to what we found, improving transparency will help in that case.	
Pearce et al. \cite{Pearce2022AsleepContributions} touch upon the fact that	GitHub Copilot is a black-box and a closed resource residing on a remote server which is unaccessible by general users.	
Cheng et al. \cite{Cheng2022ItTools} raises that getting user content is key as per GDPR privacy principles. The authors recommend the users of AI code generating tools need to get the informed consent from paricipants. This is also related to being transparent (the previous point) as in order to get user consent, the tool needs to be transparent about its actions.		
As per the research by Tian et al. \cite{Tian2023IsIt}, long prompts tend to have a negative impact on code generated by ChatGPT and Codex. Therefore, the participants need to master prompt engineering.	
Camara et al. \cite{Camara2023OnUML} found that longer details result in poor outputs.	
By going through experience reports of developers, the authors of \cite{Sarkar2022WhatIntelligence} found that the developers find it hard to write prompts.	
Cheng et al. \cite{Cheng2022ItTools} recommend that AI code generating tools have to warn users in cases like sharing information to the public.	
In their work \cite{Tang2023AnCode}, Tang et al. state the participants found that the code generated in general is similar to style and quality of a human but for a couple of the experiments the participants participated found that the code generated were not similar to the human style. As mentioned previously, they also found the generation of large chunks of code cumbersome.	
To the best of our knowledge, we could not find any particular study specifically focusing on AI-based code generation for privacy that revealed insights on the need of understanding diverse data privacy laws, self testing, explainable output, compliance 
which are imperative findings of our study.

\section{Conclusion}
AI-based programming assistants have rapidly become very popular. We conducted a survey of 51 professional developers that identified a range of using, perceived usefulness, and concerns about these tools. 
Our study revealed common usage challenges and improvements for AI-based programming assistants available in the market. AI-based programming assistants do need to be improved more for developers to trust them with generating code for ensuring data protection (privacy). We developed several recommendations for developers and AI-based programming assistants producing companies to improve them for privacy-preserving code generation tasks.

\section*{Acknowledgments}
Madampe and Grundy are supported by ARC Laureate Fellowship FL190100035. Our sincere gratitude goes to the participants who took part in the study.

\bibliographystyle{ACM-Reference-Format}
\bibliography{main}

\end{document}